# Enthalpy-based Thermal Evolution of Loops: II. Improvements to the Model

P. J. Cargill[1,2], S.J. Bradshaw[3] and J.A. Klimchuk[4]

1. Space and Atmospheric Physics, The Blackett Laboratory
2. , Imperial College, London SW7 2BW (p.cargill@imperial.ac.uk)
3. School of Mathematics and Statistics, University of St Andrews, St Andrews, Scotland KY16 9SS
4. Department of Physics and Astronomy, Rice University, Houston, TX 77005
5. NASA Goddard Space Flight Center, Solar Physics Lab., Code 671, Greenbelt, MD 20771

Abstract

This paper develops the zero-dimensional (0D) hydrodynamic coronal loop model "Enthalpy-based Thermal Evolution of Loops" (EBTEL) proposed by Klimchuk et al (2008), which studies the plasma response to evolving coronal heating, especially impulsive heating events. The basis of EBTEL is the modelling of mass exchange between the corona and transition region and chromosphere in response to heating variations, with the key parameter being the ratio of transition region to coronal radiation. We develop new models for this parameter that now include gravitational stratification and a physically motivated approach to radiative cooling. A number of examples are presented, including nanoflares in short and long loops, and a small flare. The new features in EBTEL are important for accurate tracking of, in particular, the density. The 0D results are compared to a 1D hydro code (Hydrad) with generally good agreement. EBTEL is suitable for general use as a tool for (a) quick-look results of loop evolution in response to a given heating function, (b) extensive parameter surveys and (c) situations where the modelling of hundreds or thousands of elemental loops is needed. A single run takes a few seconds on a contemporary laptop.

Final Version. April 2012

**1. Introduction**.

Since the recognition in the 1970s that the magnetically confined solar corona is comprised of discrete loops, a great deal of effort has been devoted to modelling the temporal evolution of loop plasma. One approach is to solve numerically the one-dimensional hydrodynamic (1D hydro) equations of mass, momentum and energy conservation along a magnetic field line (or strand, or loop) in response to an imposed time-dependent heating function representing a flare or smaller heating event (e.g. Peres, 2000). Of importance is the ability of such models to generate "observables" that can be used to interpret coronal data (e.g. Hansteen, 1993; Bradshaw and Cargill, 2006; Bradshaw and Klimchuk, 2011).

1D hydro models have two difficulties. One is the optically thick chromosphere at the lower boundaries. In principle this requires a full radiative-hydrodynamic treatment (e.g. McClymont and Canfield, 1983) but one can attach a simple lower atmosphere that preserves the essential physics (e.g. Klimchuk et al., 1987; Antiochos et al., 1999). The second, and more significant, difficulty is the limitation imposed on the computational timestep by thermal conduction in the transition region (hereafter TR). In static equilibrium loops (e.g. Martens, 2010) the downward heat flux implies a temperature scale height ($L_T$) of under 1 km in the TR, and even shorter in hot flaring loops. Resolving this requires a fine grid, but when modelling thermal conduction the timestep scales as the smallest value of $L_T^2$, implying long run times.

There is thus a need for simple and fast ways of modelling the coronal response to time-dependent heating. "Zero-dimensional" (0D) models, which average over the loop's spatial dimension (Kuin and Martens, 1982, Fisher & Hawley, 1990, Kopp and Poletto, 1993, Cargill, 1994, Klimchuk et al., 2008, Aschwanden and Tsiklauri, 2009) accomplish this. In addition to providing "quick look" results, 0D models are useful if a loop is comprised of many hundreds or thousands of thin, thermally isolated, randomly heated strands (Cargill, 1994), which conventional 1D hydro modelling still finds a large task. They can also provide physical insight obscured in 1D models.

The success of 1D and 0D models of this type depends on handling correctly the exchange of matter between the corona, TR and chromosphere in response to a changing coronal temperature. While the above 0D models all address this to varying degrees (see Cargill et al., 2012, hereafter Paper 3), we base our discussion here on the work of Klimchuk et al., (2008: hereafter Paper 1) where we developed a 0D model whose centrepiece was the calculation of the enthalpy flux to and from the corona. The model, called EBTEL: "Enthalpy Based Thermal Evolution of Loops", divides a loop into coronal and TR parts, the boundary being defined as where thermal conduction changes from a loss to a gain. Whether the enthalpy flux is into, or out of, the corona depends on whether the TR can radiate away the downward heat flux. If it cannot, then material is "evaporated" into the corona, whose density then increases (e.g. Antiochos and Sturrock, 1978). If the downward heat flux is too small to power the TR radiation, then there must be a downward enthalpy flux, and the coronal density decreases (e.g. Cargill et al., 1995). The top of the TR is then where enthalpy changes from a coronal loss to a TR gain. The model was compared with 1D hydro simulations of an impulsively heated loop (starting each time with the same initial conditions), and gave reasonable agreement.

EBTEL relies on three parameters, the most important of which is the ratio of the TR to coronal radiative losses. They govern both the initial equilibrium and how the loop cools after impulsive heating. It has become apparent through use of EBTEL, and attempts to benchmark the results against other known solutions of loop cooling, that the choice of this parameter in Paper 1 was not optimal for many circumstances. The physical principles behind EBTEL are unchanged, but, by re-examining the three key parameters, we here put the EBTEL model on a broader foundation. A wider range of heating events are also shown. The result is a model that, when compared with a 1D hydro code, can now follow with satisfactory accuracy the evolution of loops over a range of lengths and temperatures. In Paper 3 we will provide a comparison of 0D models and sources of potential discrepancy with 1D models.

**2. Overview of the EBTEL model**

**2.1 The governing equations**

The details of the model are discussed in Paper 1 and so are just restated briefly here. The 1D energy equation is:

$$\frac{\partial E}{\partial t} = -\frac{\partial}{\partial s}v(E+p) - \frac{\partial F_c}{\partial s} + Q - n^2 \Lambda(T) \qquad (1)$$

where $v$ is the velocity, $E = \frac{p}{\gamma - 1}$, $F_c = -\kappa_0 T^{5/2}\frac{\partial T}{\partial s}$ is the heat flux, $Q(t)$ is a heating function that includes both steady and time-dependent components, $\Lambda(T) = \chi T^\alpha$ is the radiative loss function in an optically thin plasma as defined in Paper 1, Equation (3), and $s$ is a spatial coordinate along the magnetic field. We have assumed that the flow is always subsonic and that gravity can be neglected in the energy equation. There is also an equation of state: $p = 2nk_B T$.

For a corona loop of half-length $L$ and a transition region of thickness $l$ ($<<L$), we define the boundary between corona and TR as the location where conduction changes from a loss to a gain (Vesecky et al., 1979). Integrating Eq (1) from the top of the TR to the top of the loop and enforcing symmetry boundary conditions, we find:

$$\frac{L}{\gamma-1}\frac{d\overline{p}}{dt} = \frac{\gamma}{\gamma-1}p_0 v_0 + F_0 + L\overline{Q} - R_c \qquad (2)$$

where "overbar" denotes an averaged coronal quantity, subscript "0" denotes a quantity at the base of the corona (or top of the TR) and $R_c = \overline{n}^2 \Lambda(\overline{T})L$. Note that the heat flux and enthalpy flux can play equivalent roles in providing energy to the TR.

Integrating over the TR, and assuming the heat flux and flow are small at its base, the pressure derivative and the heating can be eliminated since $l << L$, giving:

$$\frac{\gamma}{\gamma-1}p_0 v_0 + F_0 + R_{tr} = 0 \qquad (3)$$

where $R_{tr}$ is the integrated radiative TR losses. Eq (3) can then be combined with Eq (2) to give an equation for the coronal evolution:

$$\frac{1}{\gamma-1}\frac{d\overline{p}}{dt} = \overline{Q} - \frac{1}{L}(R_c + R_{tr}) \qquad (4)$$

Note that conduction and enthalpy do not appear in Eq (4) emphasising their roles as energy redistribution mechanisms in the loop as opposed to energy losses or gains. The density evolution comes from a similar approach, and in the corona we find:

$$\frac{d\overline{n}}{dt} = \frac{n_0 v_0}{L} = -\frac{\gamma-1}{2kT_0 L\gamma}(F_0 + R_{tr}). \qquad (5)$$

The average coronal temperature then follows from the equation of state:

$$\frac{1}{\overline{T}} \frac{d\overline{T}}{dt} = \frac{1}{\overline{p}} \frac{d\overline{p}}{dt} - \frac{1}{\overline{n}} \frac{d\overline{n}}{dt} \qquad (6)$$

To solve the set of coronal equations (4) – (6) for the primary variables $\overline{T}$, $\overline{n}$ and $\overline{p}$, we need to relate $R_{tr}$, $T_0$ and $F_0$ to coronal quantities. The conductive losses are defined in terms of the loop apex temperature ($T_a$): $F_0 = -(2/7)\kappa_0 T_a^{7/2}/L$ (Paper 1: Eq 20), so three temperatures characterise the corona: $\overline{T}$, $T_a$ and $T_0$. $T_a$ and $T_0$ are defined as $C_2 = \overline{T}/T_a$, $C_3 = T_0/T_a$. Finally, we define a third parameter: $C_1 = R_{tr}/R_c$ which leads to Eqs (4) and (5) becoming:

$$\frac{1}{\gamma-1} \frac{d\overline{p}}{dt} = \overline{Q} - \frac{R_c}{L}(1+C_1) \qquad (7)$$

$$\frac{d\overline{n}}{dt} = \frac{n_0 v_0}{L} = -\frac{C_2}{C_3} \frac{(\gamma-1)}{2k\overline{T}L\gamma}(F_0 + C_1 R_c) \qquad (8)$$

Eq (6), (7) and (8) can then be solved on specification of $C_{1-3}$. Initial conditions come from solving the steady state versions of (7) and (8). This approach gives slightly different values from the familiar scaling law results (e.g. Martens, 2010) since our choices of $C_1$ are approximations to the exact static loop structure.

Apex quantities are also useful when comparing with 1D hydro models. For a semi-circular loop, the apex and average pressures are $p_a \approx p_0 \exp(-2L/\lambda(\overline{T})\pi)$ and $\overline{p} \approx p_0 \exp(-2L\sin(\pi/5)/\lambda(\overline{T})\pi)$ where $p_0$ is the pressure at the top of the transition region, $\lambda(\overline{T}) = 2k\overline{T}/m_i g$ is the coronal scale height based on the average temperature and ion mass, and the factor $\sin(\pi/5)$ is discussed in Section 3.1. EBTEL calculates the average pressure, so it is straightforward to then work out $p_a$. The same is true for the density: $n_a = \overline{n} C_2 \exp(-2L(1-\sin(\pi/5))/\lambda(\overline{T})\pi)$.

**2.2 The calculation of constant $C_1$**

In Paper 1 we used constant values of $C_{1-3}$ calculated from static equilibrium loop solutions. Two approaches were considered. The first, used to produce all the Figures in Paper 1, adopted fixed values of $C_{1-3}$ at all temperatures, namely $C_1 = 4$, $C_2 = 0.87$ and $C_3 = 0.5$. We refer to these as the "EBTEL-1" values. The second used a polynomial fit for $C_1$ and $C_3$ over the temperature range 1 – 10 MK (Tables 1 and 2 of Paper 1). However the values of $C_1$ and $C_3$ in Table 1 of Paper 1 are incorrect for short loops and T > 3 MK.

Thus, the values of $C_{1-3}$ appropriate for static equilibrium loops must be reassessed. We first neglect gravity and use a simple power law radiative loss function of the form $\Lambda(T) = \chi T^{\alpha} = 1.95 \; 10^{-18} \; T^{-2/3}$ above $10^{4.97}$ K, and $\Lambda(T) = 1.1 \; 10^{-31} T^2$ below $10^{4.97}$ K to avoid unrealistic losses at low temperatures. In Appendix A the work of Martens (2010) is used to demonstrate analytically that $C_1$ and $C_3$ are independent of all loop parameters except the slope of the radiative loss function when there is no low temperature correction to $\Lambda(T)$. Then, modifying $\Lambda(T)$ at low temperatures, hydrostatic thermal equilibrium is calculated numerically. $T_a$ and $L$ are specified, and a double iteration calculates the base pressure ($p_b$) and (constant) heating subject to appropriate boundary conditions at the top of loop ($T = T_a$ and $dT/ds = 0$) for a small base temperature and vanishing base heat flux. This gives the usual scaling laws between $T_a$, $L$, $Q$ and $p_b$. These solutions show that $C_{1-3}$ may be taken as constants over a wide range of $T_a$ and $L$. For $L$ = 2.5, 5 and 7.5 $10^9$ cm and $T_a$ between 0.5 - 10 MK, $C_2$ and $C_3$ are 0.89 and 0.6 respectively. $C_1$ varies a little more with $T_a$, but can be taken as approximately 2. We propose these as the baseline values of $C_{1-3}$ in the absence of gravity and refer to them as "EBTEL-2" values.

**2.3 Comparison between 0D and 1D models**

The EBTEL results are compared with the 1-D Hydrad code (Bradshaw and Mason, 2003, Bradshaw and Cargill, 2006). Hydrad solves time-dependent electron and ion energy equations together with equations of mass and momentum conservation along a magnetic field line, and an equation of state. Here we introduce an anomalously high electron-ion collision frequency to ensure equal electron and ion temperatures. The optically thin radiative loss function is the same as in EBTEL (Paper 1, Eq (3)). In all cases EBTEL uses the same coefficient of thermal conductivity ($\kappa_0$ = 8.12 $10^{-7}$ ergs cm$^{-2}$ s$^{-1}$ K$^{-7/2}$) and average ion mass ($m_i$ = 2.17 $10^{-24}$ g) as Hydrad.

In Paper 1, we compared EBTEL results with average coronal values from the 1D ARGOS code (Antiochos et al., 1999), with the ARGOS averages calculated over the upper 80% of the loop. However, our static loop calculations show that the TR occupies at most only the lower 10% of a loop, so averaging Hydrad results over 80% will under-estimate the average coronal density. Coronal averages from Hydrad are now calculated over 90% of the loop. [It should be stressed that considerable experimentation with both ARGOS and Hydrad showed no reliable way of identifying the TR/coronal boundary in such 1D codes, in part because in the more dynamic phases there can be multiple locations where the conduction changes from a loss to a gain.] We also show some comparisons between apex quantities, with the apex results from Hydrad being spatial averages over 10% at the loop top. However, the density and pressure from Hydrad are rather spiky due to the interaction at the loop top of evaporation fronts from each footpoint, leading to a compression, and subsequent oscillation that persists for a few periods. A smoothing over roughly 30% of the oscillation period is applied to the Hydrad apex average results.

## 2.4 Results for constant $C_{1-3}$

We revisit an example from Paper 1: a long loop heated by a small nanoflare (hereafter Case 1). Table 1 provides a list of all cases, showing the heating function, loop half-length, initial temperature and density. The heating pulse is triangular with half-width ($t_H$) and magnitude ($H_0$). For Case 1 $t_H$ = 250 sec, $H_0$ = 1.5 $10^{-3}$ ergs cm$^{-3}$ s$^{-1}$ and L = 75 Mm. For a strand diameter of 200 km, 1.77 x $10^{24}$ ergs is released.

Figure 1 shows the average temperature, pressure, density, apex density, fractional errors in T and n and relationship between T and n for the "EBTEL-1" parameters. The thick (thin) solid lines are Hydrad (EBTEL-1) results, except in the error plot where the thin (thick) lines are temperature (density) and solid (dashed) lines correspond to average (apex) quantities. Hydrad apex quantities are smoothed over a 500 sec window to reduce spikiness. In the *T-n* plot (lower right panel), we show $n/n_{max}$ and $T/T(n_{max})$ where $n_{max}$ is the maximum density and $T(n_{max})$ is the temperature at the time of maximum density. Table 2 provides a summary of the results, namely the maximum temperature, density and pressure, the time these are reached, the time interval over which their values exceed 90% of the maximum, and the scaling between $T$ and $n$, assumed to be of the form $T\sim n^\delta$, in the cooling phase. As we will discuss in Section 3.2, the last of these is a very important diagnostic of loop cooling. The stars on the temperature, density and $T$-$n$ plots show the start and end times over which this scaling is evaluated. These are chosen when radiative losses are the most important coronal cooling process.

Figure 1 and the first two rows of Table 2 show the following: (i) the maximum EBTEL-1 average temperature and density exceed the Hydrad values by 12% and 8% respectively, and the average pressure combines these differences. (ii) The time of the maximum average density and temperature agree to 110 and 30 secs respectively, and the time when they exceed 90% of the maximum shows similar differences. The EBTEL pressure maximum precedes the Hydrad one by longer. (iii) In the decay phase, the EBTEL temperature (density) is systematically above (below) the Hydrad values, as can also be seen in the error plot, and (iv) this leads to a relationship between T and n in the decay phase characteristic of equilibrium as opposed to cooling loops. (v) The apex densities are in superficially better agreement, but the spikiness in the Hydrad result may exaggerate this. (vi) Both Hydrad and EBTEL return the loop to its pre-heated state after $10^4$ secs.

Figure 2 and the 3$^{rd}$ row of Table 2 show results for the "EBTEL-2" values of $C_{1-3}$. Here: (i) the EBTEL-2 maximum temperature and density increase over EBTEL-1, the temperature exceeding the Hydrad value by 20%. (ii) The time of the maximum EBTEL density is now delayed with respect to Hydrad. (iii) The EBTEL density in the decay phase is now persistently higher than Hydrad and in turn this leads to (iv) an EBTEL-2 $T$-$n$ scaling more typical of a cooling loop, but notably steeper than Hydrad. This occurs because larger (smaller) values of $C_1$ imply that the TR is more (less) efficient at radiating away downward energy fluxes (conduction or enthalpy). Larger values of $C_1$ thus lead to smaller coronal densities. For smaller values of $C_1$, the larger coronal density leads to slower conductive cooling (the conductive cooling time scales with density), hence higher peak temperatures and later times for the density maximum.

However, this pair of models shows that neither sets of constant values of $C_{1-3}$ does well at all times, and in some ways the EBTEL-2 choice makes things worse. We have identified this as being due to the absence of two pieces of physics: (i) the lack of a description how gravity changes the loop behaviour and (ii) the incorrect handling of the decay phase where the T~ $n^{1/2}$ scaling from static equilibrium models does not apply, and loops should be "over-dense" with respect to equilibrium values.

## 3. Inclusion of additional physics in EBTEL

### 3.1 Re-assessment of parameters: equilibrium loops

The first piece of missing physics to be discussed is the inclusion of gravity. The main effect is that, while the TR radiation is driven by the downward heat flux, and so for a given coronal temperature and loop length is roughly fixed, the coronal radiation falls due to stratification. Thus larger values of $C_1$ can be expected for loops with significant ratios of the length to the gravitational scale height. We have solved the hydrostatic equations for semi-circular equilibrium loops using the simple power law radiative loss function mentioned in Section 2.2 for $L = 5 \times 10^9$ and $7.5 \times 10^9$ cm, and temperatures between $5 \times 10^5$ and $4 \times 10^6$ K. In the upper panels of Figure 3, the stars denote $C_1$ when gravity is absent (around 2 in all cases) and the circles show $C_1$ when gravity is included. [Note that static solutions for $T_a = 5 \times 10^5$ K with $L = 7.5 \; 10^9$ cm could not be found: see also Serio et al., 1981.] $C_1$ increases as the temperature and scale height decrease. $C_2$ and $C_3$ have negligible dependence on gravity.

We now parameterise $C_1$ in the form $C_1(T_a, L)$. There is little dependence of $C_1$ on $L$ itself, rather the key parameter is the ratio of half-length to scale height. We write:

$$C_1 = \frac{R_{tr}}{R_c} = \left[\frac{R_{tr}}{R_{tr}(g=0)}\right]\left[\frac{R_{tr}(g=0)}{R_c(g=0)}\right]\left[\frac{R_c(g=0)}{R_c}\right] \quad (9)$$

and calculate the three ratios on the right hand side. The label "$g=0$" are values when gravity is neglected. The lower panels of Figure 3 show $R_{tr}(g=0)/R_{tr}$ (stars) and $R_c(g=0)/R_c$ (circles) as a function of $T_a$. As anticipated, the first ratio in Eq (9) is roughly unity. From Section 2.2, the middle ratio in Eq (9) is 2. Figure 3 shows that the third ratio has the expected drop when gravity is included. To calculate an approximate form for $C_1$ we argue that, for a given coronal temperature,

$$\frac{R_c(g=0)}{R_c} \approx \frac{\left(\bar{n}^2 \bar{T}^\alpha\right)_{g=0}}{\bar{n}^2 \bar{T}^\alpha} \approx \frac{\left(\bar{n}^2\right)_{g=0}}{\bar{n}^2} \approx \frac{p_0^2}{\bar{p}^2} \qquad (10)$$

assuming the coronal half-length is the same with and without gravity.

Next assume that for a semi-circular loop the coronal pressure is given by: $p(s) \approx p_0 \exp\left[-\frac{2L}{\pi \bar{\lambda}} \sin\left(\frac{\pi s}{2L}\right)\right]$ where we use a scale height based on the average temperature ($\bar{T} = C_2 T_a$) and that hydrostatic density stratification occurs only in the corona. Integrating $p(s)$ numerically gives an average pressure, and this average value is well approximated by using the actual pressure at $s/L = 0.4$. So the average pressure is written as: $\bar{p} = p_0 \exp(-2L\sin(\pi/5)/\bar{\lambda}\pi)$ (which also accounts for the $sin(\pi/5)$ factor introduced in Section 2.1) and so:

$$C_1 = \frac{R_{tr}}{R_c} = 2\exp(4L\sin(\pi/5)/\bar{\lambda}\pi) \qquad (11)$$

The plus signs in the upper panels of Fig 3 show that Eq (11) works well for all but the lowest temperature.

An analysis can also be carried out to include multiple power law loss functions. Since the method is similar, it is discussed in Appendix B. One can finally obtain a formula for $C_1 = C_1(eqm)$ for equilibrium loops including gravity and radiation as:

$$C_1(eqm) = 2\left[\frac{R_c(g=0, \alpha=-2/3)}{R_c(g=0,\bar{T})}\right]\exp(4L\sin(\pi/5)/\bar{\lambda}\pi) \qquad (12)$$

where the ratio is that of the radiative losses for the -2/3 power law to that calculated using the full EBTEL loss function.

**3.2 Radiative cooling phase**

In a static equilibrium loop, there is a scaling between temperature and density of order $T \sim n^{1/2}$ that arises because coronal conductive and radiative losses are roughly equal. However, this does not hold in the cooling phase after the density maximum when the energetics involve mainly coronal cooling due to radiation and an enthalpy flux to the TR. For short, hot loops, there is a scaling $T \sim n^2$ during this phase (Serio et al., 1991; Cargill et al., 1995; Bradshaw and Cargill, 2005, 2010a,b), with a scaling approaching $T \sim n$ for longer, more tenuous loops (Bradshaw and Cargill, 2010b). We can adapt the EBTEL equations to this regime and determine the appropriate value of $C_1$. Neglecting thermal conduction and heating, Eq (7) and (8) are:

$$\frac{1}{\gamma-1}\frac{d\bar{p}}{dt} = -\frac{R_c}{L}(1+C_1), \quad \frac{d\bar{n}}{dt} = -\frac{C_2 C_1}{C_3}\frac{(\gamma-1)R_c}{2k_B \bar{T} L \gamma} \quad (13)$$

and so, on writing $T \sim n^\delta$, we can relate $T$ and $n$:

$$\delta + 1 = \frac{1}{p}\frac{dp}{dt} \div \frac{1}{n}\frac{dn}{dt} = \gamma \frac{C_3}{C_2}\frac{(1+C_1)}{C_1} \quad (14)$$

This can be solved for $C_1$ as:

$$C_1(rad) = \left(\frac{C_2(1+\delta)}{\gamma C_3} - 1\right)^{-1} \quad (15)$$

where $T_0$ is now the temperature at which enthalpy changes from a loss to a gain (Bradshaw and Cargill, 2010a,b) and we denote $C_1(rad)$ as $C_1$ in the radiative phase. For $\delta = 2$ (1), and the same values of $C_2$ and $C_3$ derived above, $C_1 = 0.6$ (1.25). We chose $C_1 = 0.6$ as the baseline value for the radiative phase in the absence of gravity and a loss function coefficient of $\alpha = -2/3$.

To include gravity and a full $\Lambda(T)$, we adopt the same approach as in Section 3.1 and Appendix B based on our work on radiative cooling (Cargill et al., 1995; Bradshaw and Cargill, 2005, 2010a,b), with gravity increasing $C_1(rad)$:

$$C_1(rad) = \frac{R_{tr}}{R_c} = 0.6\left[\frac{R_c(g=0, \alpha=-2/3)}{R_c(g=0, \bar{T})}\right]\exp(4L\sin(\pi/5)/\bar{\lambda}\pi). \quad (16)$$

Eq (16) is equivalent to Eq (12) except that the coefficient "2" is replaced by "0.6".

### 3.3 Overall implementation of $C_1$

We now implement a formalism for $C_1$ that has a smooth transition from equilibrium to radiative values as the loop evolves after the density maximum. The formalism is expressed in terms of the ratio of the actual average density to the density for a loop in static equilibrium at a given temperature. This is a measure of the "over-density" of the loop, and it is well known that loops undergoing radiative cooling are in this regime (e.g. Cargill and Klimchuk, 2004). For a given average temperature, Eq (7) and (8) give the density required for equilibrium as:

$$n_{eq}^2 = \frac{2\kappa_0}{7C_1(eqm)\Lambda(\overline{T})L^2}\left(\frac{\overline{T}}{C_2}\right)^{7/2} \qquad (17)$$

Now $(n/n_{eq})^2 = C_1(eqm)\tau_c/\tau_R$ where $\tau_c$ and $\tau_R$ are the conductive and radiative cooling times, defined as: $\tau_c = \frac{21k_B\overline{n}L^2C_2}{2\kappa_0}\left(\frac{C_2}{\overline{T}}\right)^{5/2}$, $\tau_R = \frac{3k_B\overline{T}}{\overline{n}\Lambda(\overline{T})}$. So radiative-dominated cooling of an impulsively heated loop at a given temperature is characterized by $n > n_{eq}$ (e.g. Cargill, 1994, Cargill and Klimchuk, 2004). We thus define $C_1$ as:

$$C_1 = C_1(eqm) \qquad n < n_{eq}$$

$$C_1 = \left[2C_1(eqm) + C_1(rad)((n/n_{eq})^2 - 1)\right]/\left[1 + (n/n_{eq})^2\right] \qquad n > n_{eq} \qquad (18)$$

which is piecewise continuous at $n = n_{eq}$.

### 3.4 Case 1 revisited

We return to Case 1 and use Eq (18) for $C_1$. $C_2$ and $C_3$ are unchanged. Figure 4 has the same format as Figure 1 and, along with row 4 of Table 2, shows that (i) the EBTEL and Hydrad maximum temperatures still differ by roughly 20%. (ii) While the EBTEL density still is larger at all times than the Hydrad values, the difference in the average density is diminished (the difference at the maximum is now under 10%), and the apex density shows very good agreement. (iii) The delay in the timing of the maximum density present in the "EBTEL-2" run has been largely removed and (iv) the *T-n* scaling in the decay phase is now closer to the Hydrad value, and characteristic of a loop cooling by radiation.

The physics behind this involves a number of factors and we have turned on and off various terms in the $C_1$ parameterisation to clarify what is going on. First, retaining the gravitational physics in $C_1$ and ignoring the radiative phase correction maintains the value of the peak density but leads to decay phase densities that are too low and a *T-n* scaling of $T \sim n^{0.5}$, typical of an equilibrium loop. Secondly, keeping the radiative correction and ignoring gravity gives high densities at all times (by almost a factor of two in the decay phase) and a *T-n* scaling of $T \sim n^{2.4}$. Thus including both the radiative physics and stratification are essential. Ignore either, and the decay phase is not modelled properly. Ignore gravity, and the density is too high.

**4. Further Results**

Case 1 has demonstrated explicitly how the inclusion of gravitational and radiative decay physics can enhance the performance of EBTEL. Case 1 is quite challenging because of the extreme loop length, low density, and consequent strong role of gravity. We now present three more cases that each pose specific challenges. Case 2 is a more typical coronal problem, a nanoflare in a medium length loop, and poses a challenge to both the gravitational and radiative physics. Case 3 is a small flare that provides a test for the radiative physics alone since gravity is not important. Case 4 is similar to Case 2, except the nanoflare is in a loop with significantly higher density. Here we wish to see whether the loop returns to its pre-event state.

Case 2 is a nanoflare in a short loop of half-length 25 Mm. The pulse half-width is 100 sec and peak magnitude $10^{-2}$ ergs cm$^{-3}$ s$^{-1}$. For a 200 km diameter strand, 1.57 x $10^{24}$ ergs is released. Figure 5 is in the same format as the others with the addition of results using the "EBTEL-1" parameters (the dashed lines in the top four and lower right panels). The smoothing in the apex quantities is now over 150 secs. The EBTEL-1 parameters do poorly, in particular with the density, peaking too early and falling off far too rapidly. However, it is clear that the new EBTEL and Hydrad show good agreement and rows 5 and 6 of Table 2 show that the peak temperature and density differ 14% for the temperature and 7% for the density, though the EBTEL density and pressure do peak early. The *T-n* slope in the decay phase differs by 0.1, with Hydrad showing the steeper slope, but with values commensurate with enthalpy-dominated radiative cooling. It should also be noted that the return to equilibrium after the heating event shows some disagreement (see temperature results in the error plot) with the Hydrad temperature undershooting. This is discussed further in Case 3.

Case 3 is a modest flare in a short loop with $L$ = 25 Mm, a maximum heating of 2 ergs cm$^{-3}$ s$^{-1}$, and a pulse half-width 100 secs. The total heating per unit area is 5 $10^{11}$ ergs

cm$^{-2}$ which for a loop diameter 20% of the half-length gives a total energy input of 4 10$^{29}$ ergs. We neglect "thick target" heating which was discussed in Paper 1 and remains part of the EBTEL code. Figure 6 and rows 7 and 8 of Table 2 show the results in the same format as Figure 5. Once again, the "EBTEL-1" values of density do very badly. It is also noticeable now that the agreement between the peak density is superficially not as good as the previous example (14% difference), whereas the peak temperature is better (6% difference). The time of both peaks shows differences of 10 – 20 secs between Hydrad and EBTEL. While the *T-n* slopes in the radiative phase compare well, and take on typical flare values, the EBTEL density is systematically larger by 20%. Further, the Hydrad run appears to undergo a catastrophic loop cooling from about 1200 secs and we have truncated the error plot so that the results from the heating and initial phases remain clear. This radiative collapse appears to be typical in the latter phases of flare cooling (F. Reale, private communication, 2012), and some evidence was also apparent in Case 2, though this does not happen in all cooling loops (see Figure 1 - 4 of Bradshaw and Cargill, 2010b). The topic will be discussed further in another publication.

The previous examples have a nanoflare or flare with energy much larger than the background thermal energy in the loop or, equivalently, the background heating is much smaller than the peak nanoflare heating. However, nanoflare heating is not necessarily confined to a single heating / cooling cycle in a loop. Evidence now suggests that impulsive heating on occasions may be occurring in loops or strands that have not undergone such a full evaporation – draining cycle (e.g. Warren et al, 2011), so that the heating takes place in a higher ambient density.

Case 4 is a re-run of Case 2 with a higher background heating. There is one difficulty that should be noted. The initial conditions generated by Hydrad and EBTEL are not identical, a reflection of the fact that modelling the full spatial structure of a loop will give a different value of the average temperature and density from an approximate model. So long as the flare or nanoflare is "large" in the sense that at it's peak the heating is much stronger than the background, this does not matter. For Case 4 we have adjusted the background heating to give the same initial density in Hydrad and EBTEL of 9.2 x 10$^8$ cm$^{-3}$. This leads to the Hydrad (EBTEL) initial temperatures being 1.3 (1.6) MK. The choice of fixing the initial density is arbitrary, but reflects our experience that density behaviour is a sterner test of 0D models than temperature.

The results in Figure 7 and rows 9 and 10 of Table 2 need to be considered with this in mind. Not surprisingly, the EBTEL maximum temperature exceeds the Hydrad one, though by less than the discrepancy at the start. However, the densities agree well, though the EBTEL one peaks earlier by 200 secs. In particular, we note that with both EBTEL and Hydrad the loop returns to it pre-nanoflare state after a few thousand seconds.

Looking at the results overall, we can draw some general conclusions:

(i) EBTEL calculates the time of the temperature maximum and the duration when it is over 90% of the maximum to within 50 secs in all cases, but over-estimates the value of the maximum temperature by between 10 and 20%. The error is largest for weak heating in a long loop, and smallest for the small flare. The maximum temperature is determined by how well the corona can conduct heat away, and these results suggest that the simple approximation for $F_c$ gives values that are too small. Given the simplicity of our approximation to the strongly non-linear conductive losses, we regard a 10 – 20% over-estimation as satisfactory.

(ii) EBTEL over-estimates the maximum density, though by under 14% in all cases. For the long loop, the timing of the maximum density in EBTEL is delayed by 390 secs, but the start of the 90% envelope is only off by 10 secs, suggesting that the oscillation in the Hydrad results gives an "early" peak value. In the other three cases, the difference in timing is under 100 secs. One possible reason for the larger EBTEL density is the assumption of a fixed loop length in EBTEL, and the use of the 90% spatial average in Hydrad. It is well known that when a loop undergoes impulsive heating, the top of the chromosphere is pushed downward, leading to a slightly longer loop. Given that the same amount of chromospheric plasma will be heated, and fill the (longer) coronal volume, this effect will lead to a lower average density, as is seen in the Hydrad results.

(iii) In the decay phase, the Hydrad density remains lower than the EBTEL one. However, the important *T-n* relationship that describes the radiative/enthalpy cooling show good agreement. We note that these slopes are somewhat sensitive to the start and termination of the analysis windows.

(iv) While we did not document the properties of the apex density in Table 2, examination of the Figures show that the agreement between the EBTEL and Hydrad values is excellent.

(v) More negatively, the discrepancy between the EBTEL and Hydrad pressures is larger, in excess of 15% in some cases. This is partly due to the issue discussed in point (ii) above, but also suggests that the chromosphere and TR may be more efficient radiators during the early phase than our model for $C_1$ assumes.

(iv) It is interesting to note that the discrepancies between the 0D and 1D models are small enough that the 0D model may be used as a suitable proxy for 1D given that parameters determined from inversion of spectroscopic data are probably not constrained by any less that our discrepency (Judge et al., 1997; Judge, 2010).

## 5. The Differential Emission Measure

In Paper 1 we calculated separate coronal and transition region differential emission measures (DEMs), the latter by two distinct methods. The DEM is defined as: $DEM(T) = n^2 (\partial T/\partial s)^{-1}$. The modifications to EBTEL do not change the way the coronal DEM is calculated since the coronal parameters are our primary variables. On the other hand, the TR DEM relied on an assumption of constant pressure in the loop, which the introduction of gravity will invalidate. In Paper 1 we calculate the TR DEM by solving the following quadratic equation for $\partial T/\partial s$:

$$\kappa_0 T^{3/2} \left(\frac{\partial T}{\partial s}\right)^2 - 5kJ_0 \left(\frac{\partial T}{\partial s}\right) - \left(\frac{p_{TR}}{2k_B T}\right)^2 \Lambda(T) = 0 \quad (19)$$

While $J_0 = n_0 v_0$ is determined by the mass flow to and from the corona, the pressure in the last term is a TR quantity. Thus, when gravity is important we need to modify this term to account for the fact that the TR pressure will be larger than the coronal one.

This is done by using our coronal pressure modification in reverse, so we write $p_{TR} = \bar{p}\exp(2L\sin(\pi/5)/\bar{\lambda}\pi)$. In the Appendix of Paper 1 we also provided approximate forms of the DEM for three cases of loop evolution: strong conduction-driven evaporation, equilibrium, and strong radiative-driven condensation (draining). Of these, the third is unmodified, while the first two both involve the TR pressure, and need to be changed.

The top two panels of Figure 8 show on the left the DEM from EBTEL and Hydrad for the flare Case 3 (thin and thick solid lines respectively) and the DEM for the "EBTEL-1" values of $C_{1-3}$ (dashed line) where the DEM is summed over 2000 secs. The principle difference between the EBTEL-1 values and the full model is below $10^7$ K, when the low EBTEL-1 density in the decay phase is most evident (Figure 6). The larger discrepancy at lower temperatures is due to the catastrophic loop draining in Hydrad after 1500 secs discussed in the previous Section. The top right panel shows the EBTEL DEM broken into its components from the corona (dashed) and TR (dotted). The coronal component has a slope of $T^2$ above $10^{6.2}$ and $T^{3/4}$ below that temperature. Cargill (1994, p. 387) noted that the DEM slope in this cooling phase depends solely on the slope of the radiative loss function, with a scaling of DEM ~ $T^{(1/2-\alpha)}$. The break in the slope occurs near the break in the loss function, and the slopes above and below are in general agreement with this simple scaling.

The lower left and right panels show, respectively, the DEM of a nanoflare in the long and short loop with the DEM summed over 10000 and 5000 secs respectively. While there are differences in the magnitude of the different models, the topology, which is very important for inferring coronal properties, is comparable. Here we see a flatter coronal DEM distribution below the peak compared to the flare case. This reflects the difference in the radiative cooling physics when gravity is important with a shallower *T-n* scaling. The arguments of Cargill (1994) now suggest a DEM slope of $T^{-\alpha}$, in broad agreement with what is seen.

**6. Discussion and Conclusions.**

Simple 0D hydrodynamic models have a long history in modelling the temporal evolution of transiently-heated coronal loops and in this paper we have updated our original version of the EBTEL model to include gravitational stratification and the correct radiative cooling physics. Comparison with results from the 1D Hydrad code suggest that these changes are quantifiable improvements to the original model, as can be seen by especially comparing the density of the new and original versions. It would certainly be feasible to develop further the parameterisations of $C_{1-3}$ to include more physics than we have included here, but, barring some major new understanding of how impulsively heated loops evolve, at some point diminishing returns will set in.

The applications of EBTEL were discussed extensively in the discussion of Paper 1, so only a brief summary is appropriate here. EBTEL is a useful tool in looking at the generic evolution of temperature and density, as well as the DEM of single loops. It runs fast (a few seconds on a contemporary laptop), and the output can be convolved with other software to generate, for example, light curves in various coronal emission lines. It can also be used to survey very large areas of parameter space (heating magnitude, cadence, loop length, pre-event conditions) quickly, so giving users of 1D models guidance on what to look for. But, perhaps more useful is the ability to model a multi-strand corona. In such a scenario (e.g. Cargill, 1994; Cargill and Klimchuk, 1997), the coronal emission comes from many (perhaps thousands) of separately evolving strands. This is still beyond the abilities of 1D hydro codes, at least with a realistic turn-around time whereas EBTEL can model such a scenario in a few hours, and indeed perhaps less if a properly optimised version is used.

**Acknowledgements**
JAK and SJB thank the NASA Supporting Research and Technology Program. We are grateful to the International Space Science Institute (ISSI) for supporting our team, to Helen Mason for acting as co-leader of this team with SJB and to Fabio Reale for useful discussions about catastrophic loop cooling.

# Appendix A. Comparison of analytical and numerical values of $C_1$ for a simple radiative loss function

$C_1$ and $C_3$ can be calculated analytically from Martens (2010). Assuming uniform heating, a single power law slope of $\alpha$ for the loss function, and boundary conditions of vanishing heat flux at top and bottom of loop, and vanishing temperature at bottom, he writes the energy equation in terms of the variable $\eta = (T/T_a)^{7/2}$ as:

$$\varepsilon \frac{d^2\eta}{ds^2} = \eta^\mu - \xi, \quad \mu = -\frac{2(2-\alpha)}{7}, \quad \xi = \frac{7}{3+2\alpha}, \tag{A1}$$

$$\varepsilon = \frac{2}{(1-2(2-\alpha)/7)}\left[\frac{2(2-\alpha)}{7}\right]^2 \frac{1}{B^2(\lambda+1,1/2)}, \quad \lambda = \frac{3/2+\alpha}{2(2-\alpha)}$$

where $T_a$ is the apex temperature and the scaling laws are used to eliminate $L$, $Q$ and $p$. $B(a,b)$ is a beta function. He solves the energy equation for a variable $u = \eta^{-\mu}$ as:

$$s/L = \beta_r(u, \lambda+1, 1/2),$$

where $\beta_r$ is the normalised incomplete beta function.

At the point where conduction changes from a gain to a loss, denoted by subscript zero, (A1) gives $\eta_0 = \left(\frac{7}{3+2\alpha}\right)^{-\frac{7}{2(2-\alpha)}}$ or, in real units, $\frac{T_0}{T_a} = \left(\frac{7}{3+2\alpha}\right)^{-\frac{1}{(2-\alpha)}}$.

Setting $\alpha = -1/2$, we get $C_3 = T_0/T_a = (2/7)^{2/5} = 0.606$. For $\alpha = -2/3$, $C_3 = 0.584$.

We can also calculate $C_1$ as follows. The dimensionless coronal radiative losses are:

$$\int_{s(\eta_0)}^{1} \eta^\mu ds = \xi(1 - s(\eta_0)) - \varepsilon\left(\frac{d\eta}{ds}\right)_{s=s(\eta_0)}$$

Now Eq (A1) integrates once to give, on application of the boundary conditions:

$$\frac{\varepsilon}{2}\left(\frac{d\eta}{ds}\right)^2 = \frac{\eta^{\mu+1}}{\mu+1} - \xi\eta \tag{A2}$$

so that

$$\left(\frac{d\eta}{ds}\right)_{s=s(\eta_0)} = \sqrt{\frac{2}{\varepsilon}\left(\frac{\eta_0^{\mu+1}}{\mu+1} - \xi\eta_0\right)} \tag{A3}$$

The total radiative loss is just $\xi$ in these units so that the TR loss is then:

$$\int_0^{s(\eta_0)} \eta^\mu ds = \xi s(\eta_0) + \varepsilon \left(\frac{d\eta}{ds}\right)_{s=s(\eta_0)}$$

We have calculated $\eta_0$ above, and so can obtain $C_1$, which is independent of $Q$, $L$ and $p$. For $\alpha = -1/2$, we get $C_1 = 1.76$ and for $\alpha = -2/3$, $C_1 = 2.095$.

We now compare the Martens solutions with a numerical solution that has a lower boundary at $2 \times 10^4$ K, a single power law loss function above $10^5$ K and a loss function scaling as $T^2$ below. [This eliminates the problem that the vanishing heat flux is only exactly enforceable in the limit of vanishing base temperature.] We use the following spatial grid with 5000 points:

$$s/L = (2/\pi)\left[\sin^{-1} x - x\sqrt{1-x^2}\right]$$

and $x$ is evenly distributed between 0 and 1. The motivation can be seen in Eq (C1) of Rosner et al (1978) and it does give well-resolved solutions at all temperatures.

An array of cases has been run: three loop half-lengths, 2.5, 5 and $7.5 \times 10^9$, and $T_a$ between $10^6$ and $10^7$ for each length. It turns out that the results are by and large independent of the loop half-length, so individual results are not shown, rather the ranges of values are given in Table A1. It can be seen that $C_2 = 0.89$ and $C_3 = 0.6$ are reasonable for both cases. The lower values of $C_1$ correspond to smaller $T_a$ where the $T^2$ loss function at lower temperatures makes a greater relative contribution to the loop losses. We would argue that for a simple model, $C_1 = 1.7$ for $\alpha = -1/2$ and $C_1 = 2$ for $\alpha = -2/3$ are appropriate.

**Appendix B. $C_1$ for multiple power law radiative loss function**

Neglecting gravity, we evaluate $C_1$ for a more complicated loss function by comparing results for the EBTEL loss function and the single power law one, using a similar approach to including gravity in Section 3.1:

$$C_1 = \frac{R_{tr}}{R_c} = \left[\frac{R_{tr}(T)}{R_{tr}(\alpha = -2/3)}\right]\left[\frac{R_{tr}(\alpha = -2/3)}{R_c(\alpha = -2/3)}\right]\left[\frac{R_c(\alpha = -2/3)}{R_c(T)}\right] \quad \text{(B1)}$$

where $R_{tr}(T)$ and $R_c(T)$ evaluate the loss functions at a given temperature using the full power law in EBTEL. The right hand plot in Figure 9 shows little difference in the TR losses between the two radiative loss models (stars), so we can assume the first term in (B1) is unity. The explanation is once again that the TR losses are determined by the heat flux from the corona. The coronal loss (circles) does show differences between the models. The second term in (B1) is 2. For the third term, we use the average coronal temperature ($T = \bar{T} = C_2 T_a$) in Eq (B1). The left hand plot of Figure 9 shows the same quantities as the upper left plot of Figure 3 for a loop of length 5 $10^9$ cm. This model for $C_1$ is almost independent of the loop length.

We can also combine the two corrections for loops with gravity and the general EBTEL loss function by replacing the ratio before the exponential in Eq (11) (which has gravity and the simple loss function) with Eq (B1) (which has no gravity and the full loss function), and using the fact that the TR losses are roughly the same for all cases:

$$C_1(eqm) = \frac{R_{tr}(g,\bar{T})}{R_c(g,\bar{T})} = \left[\frac{R_{tr}(g=0,\alpha=-2/3)}{R_c(g=0,\alpha=-2/3)}\right]\left[\frac{R_c(g=0,\alpha=-2/3)}{R_c(g=0,\bar{T})}\right]\exp(4L\sin(\pi/5)/\bar{\lambda}\pi)$$
(B2)

where we now denote the "equilibrium" value of $C_1$ as $C_1(eqm)$. The first ratio is 2 in this paper, but is written in a general form to allow for changes to the coronal losses that no longer use our power law approximation. Figure 10 shows the results in the same format as Figure 9 for two temperature ranges and a loop length of 5 $10^9$ cm.

| Case | $L$ ($10^9$ cm) | $H_0$ (ergs cm$^{-3}$ s$^{-1}$) | $t_H$ (s) | $T(t=0)$ (MK) | $n(t=0)$ ($10^8$ cm$^{-3}$) |
|---|---|---|---|---|---|
| 1 | 7.5 | 1.5 10$^{-3}$ | 250 | 0.85 | 0.36 |
| 2 | 2.5 | 10$^{-2}$ | 100 | 0.78 | 1.85 |
| 3 | 2.5 | 2 | 100 | 2.1 | 18.5 |
| 4 | 2.5 | 10$^{-2}$ | 100 | 1.3(1.6$^*$) | 9.2 |

**Table 1.** Summary of the cases considered. The columns show: loop half-length, maximum amplitude of triangular heating pulse, half-width of the pulse, initial average temperature and density. In case 4, the starred temperature is for EBTEL.

| $\alpha = -1/2$. | $C_1$ | $C_2$ | $C_3$ |
|---|---|---|---|
| Analytic | 1.76 | 0.89 | 0.606 |
| Numerical | 1.65 – 1.74 | 0.895 | 0.62 – 0.61 |
| $\alpha = -2/3$ | | | |
| Analytic | 2.09 | 0.89 | 0.585 |
| Numerical | 1.88 – 2.06 | 0.892 | 0.61 – 0.59 |

**Table A1**. The constants $C_1$, $C_2$ and $C_3$ for two loss functions. The range of values in each box are those obtained as $T_a$ increases from low to high.

| Case | $T_{max}(MK)$ | $t(T_{max})$ | $n_{max}(10^9)$ | $t(n_{max})$ | $p_{max}(cgs)$ | $t(p_{max})$ | $\delta$ |
|---|---|---|---|---|---|---|---|
| 1(Hydrad) | 3.94 | 260 (190-460) | 0.37 | 1450 (1020-2690) | 0.22 | 660 (460-890) | 0.83 |
| 1($C_1$=4) | 4.41 | 290 (220-430) | 0.40 | 1560 (900-2440) | 0.26 | 500 (390-1300) | 0.61 |
| 1($C_1$=2) | 4.76 | 280 (200-420) | 0.41 | 2210 (1220-3440) | 0.26 | 500 (390-1670) | 1.24 |
| 1 | 4.77 | 280 (200-420) | 0.39 | 1840 (1030-3270) | 0.26 | 500 (390-1420) | 1.02 |
| 2(Hydrad) | 3.77 | 110 (90-180) | 1.07 | 820 (430-1290) | 0.63 | 260 (190-340) | 1.33 |
| 2 | 4.30 | 120 (90-180) | 1.15 | 720 (410-1360) | 0.71 | 200 (160-580) | 1.22 |
| 3(Hydrad) | 18.9 | 120 (90 – 170) | 33.9 | 430 (260-740) | 112 | 180 (150-270) | 1.77 |
| 3 | 20.0 | 110 (80-170) | 38.7 | 450 (280-770) | 132 | 200 (150-340) | 1.89 |
| 4(Hydrad) | 3.07 | 160 (120-220) | 1.55 | 850 (390-1470) | 0.90 | 260 (150-520) | 1.22 |
| 4 | 3.44 | 170 (120-250) | 1.64 | 790 (400-1580) | 1.10 | 240 (150-730) | 1.35 |

**Table 2.** Summary of key output for the four cases shown in Table 1. The maximum of the average temperature, density and pressure are shown in columns 2, 4 and 6 and the time this maximum is reached is the upper number in columns 3, 5 and 7. The lower pair of numbers in columns 3, 5 and 7 is the time interval between which the relevant variable lies above 90% of the maximum value. All times are in seconds and have been rounded to the nearest 10 secs. In the last column, $\delta$ is defined by the relationship in the radiative cooling phase, $T \sim n^\delta$, and is calculated between the starred location on the relevant figures.

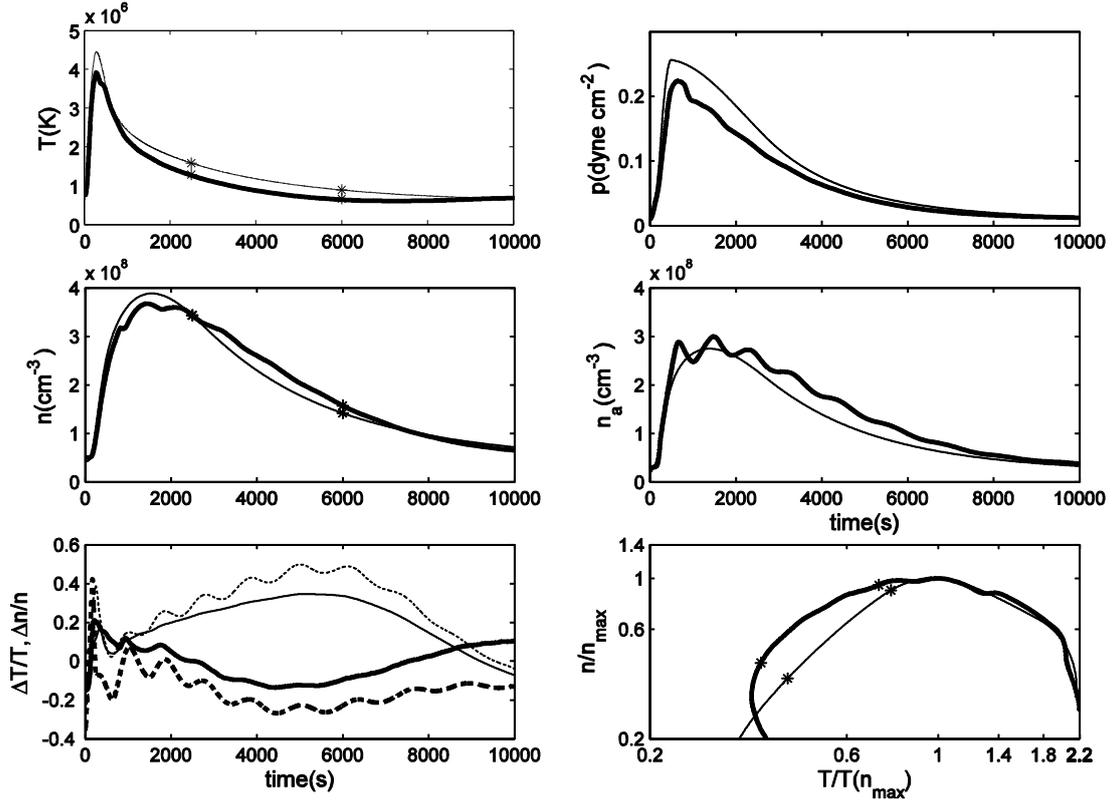

**Figure 1.** EBTEL and Hydrad solutions for a small nanoflare in a long loop (Case 1). "EBTEL-1" values of the parameters $C_{1-3}$ are used. The top four panels show the average temperature and pressure and the average and apex density. The lower right panel shows the relationship between $T$ and $n$ where $n$ and $T$ are normalised with respect to the maximum density and temperature at the time of maximum density respectively. Thick and thin solid lines are Hydrad and EBTEL results respectively. The stars show the start and end points between which the decay phase $T$-$n$ scaling is calculated. The lower left panel shows the fractional errors of the average (solid) and apex (dashed) density and temperature. In this panel the thin and thick lines correspond to the error in temperature and density respectively. The error $\Delta T/T$ is defined as [T(EBTEL) – T(Hydrad)]/T(Hydrad).

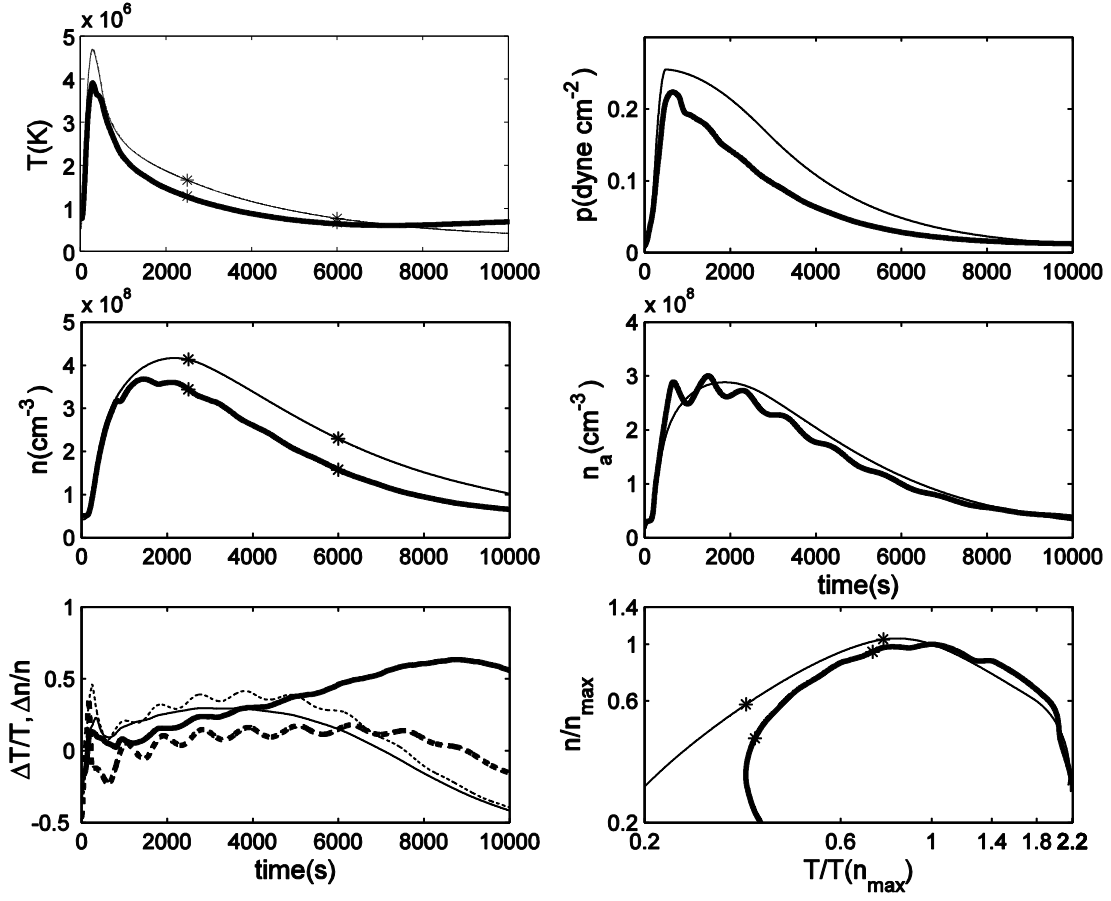

**Figure 2.** As Figure 1 except constant "EBTEL-2" values of $C_{1-3}$ are used.

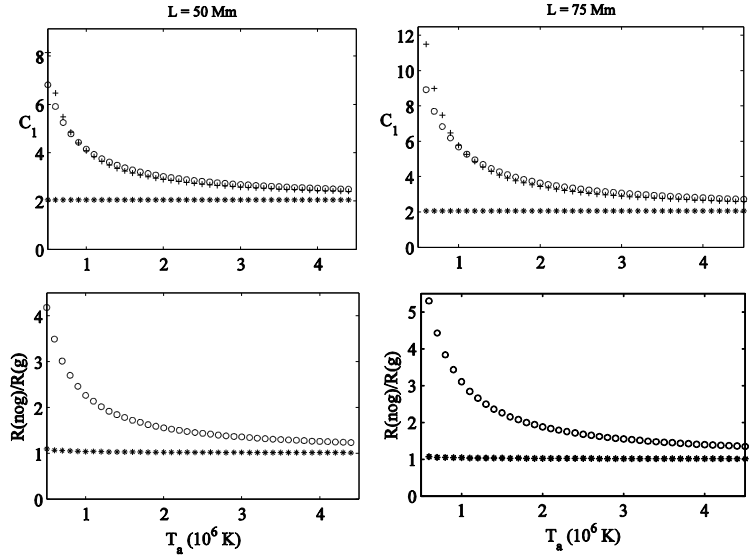

**Figure 3**: The upper two panels show the parameter $C_1$ as a function of $T_a$ for $L = 5 \cdot 10^9$ cm (left) and $7.5 \cdot 10^9$ cm (right) for a single power law loss function with a low temperature correction. Stars, circles and plus signs are, respectively, $C_1$ in absence of gravity, $C_1$ with gravity (both are from numerical solutions of the hydrostatic

equations) and the estimate of $C_1$ in Eq (11). The lower two panels show the ratio of the radiative losses without gravity to those with gravity in the transition region (stars) and corona (circles). The ratio of the two transition region losses is roughly constant.

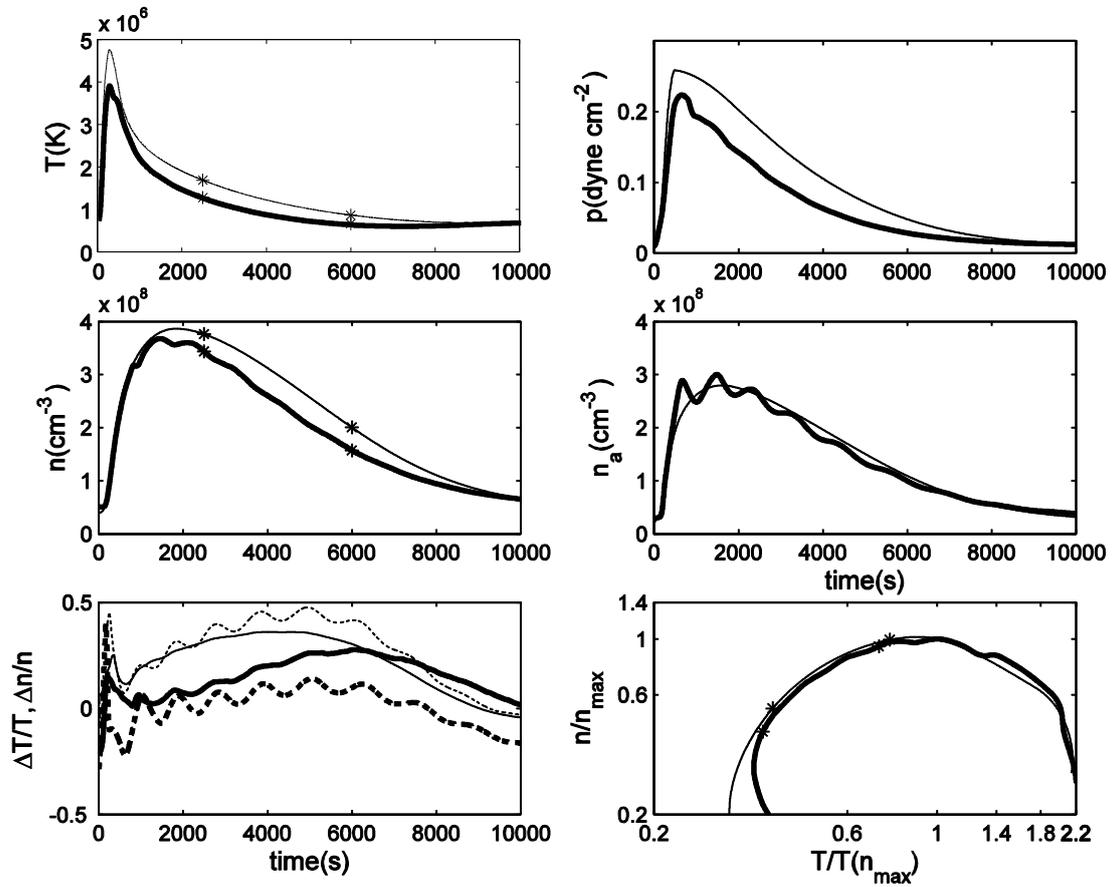

**Figure 4**: Case 1. As Figure 1, with variable $C_1$ from Eq (18) included

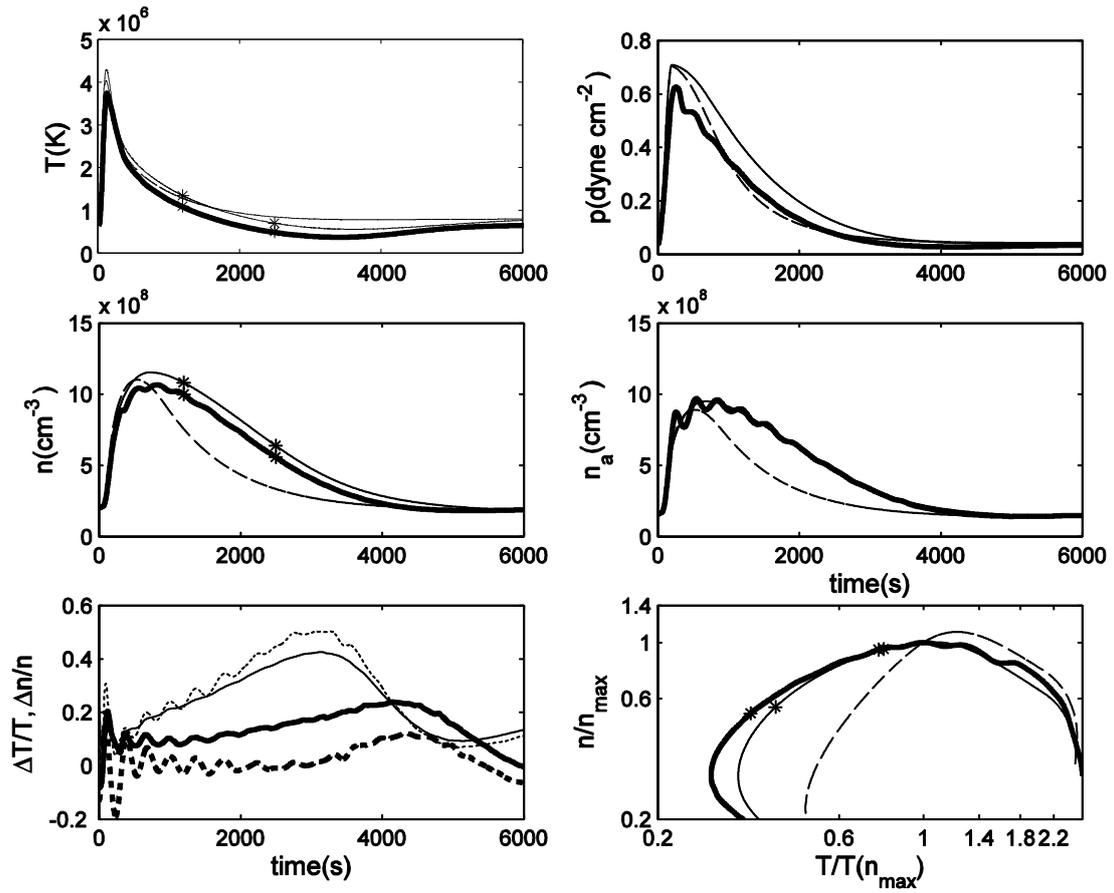

**Figure 5**. Case 2: a short loop (25 Mm half-length) with a nanoflare energy release. The format is the same as Figure 1 except that the dashed line in the top four and lower right panels is the result for EBTEL-1 parameters.

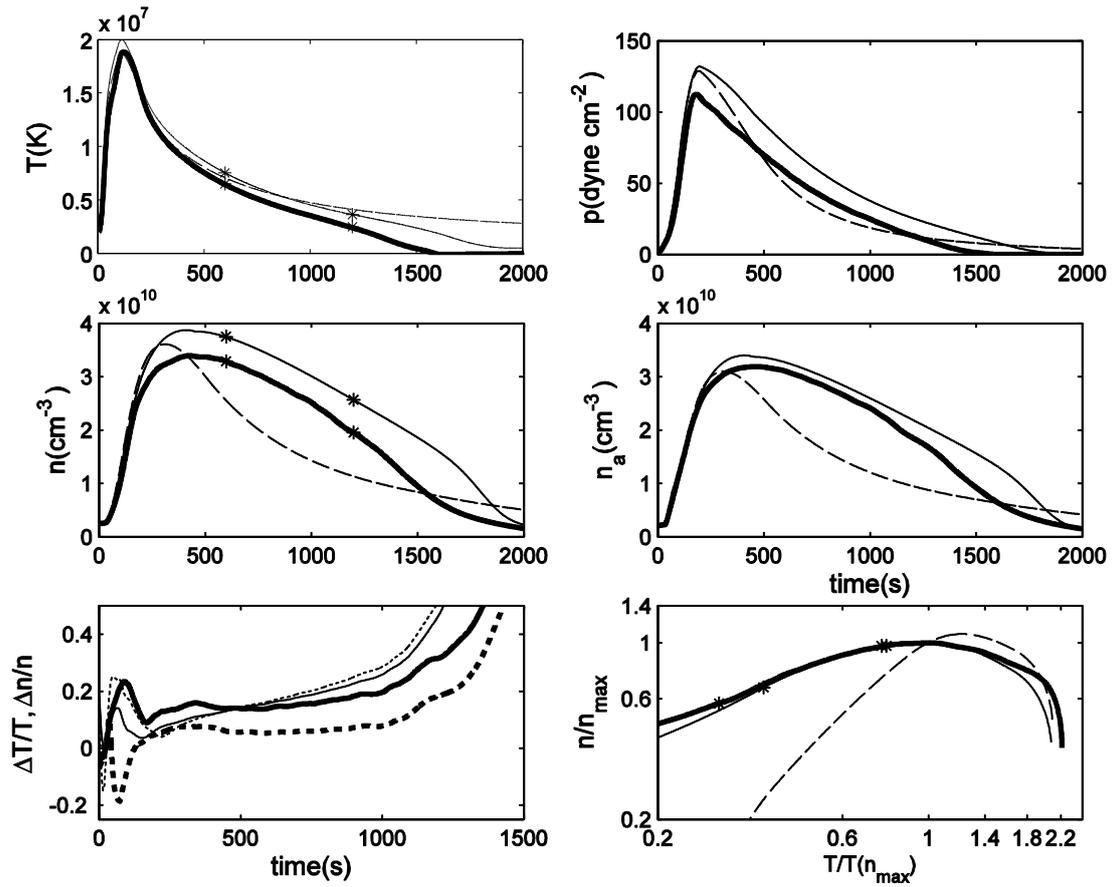

**Figure 6** Case 3: a small flare in a loop of half-length 25 Mm. The format is the same as Figure 5.

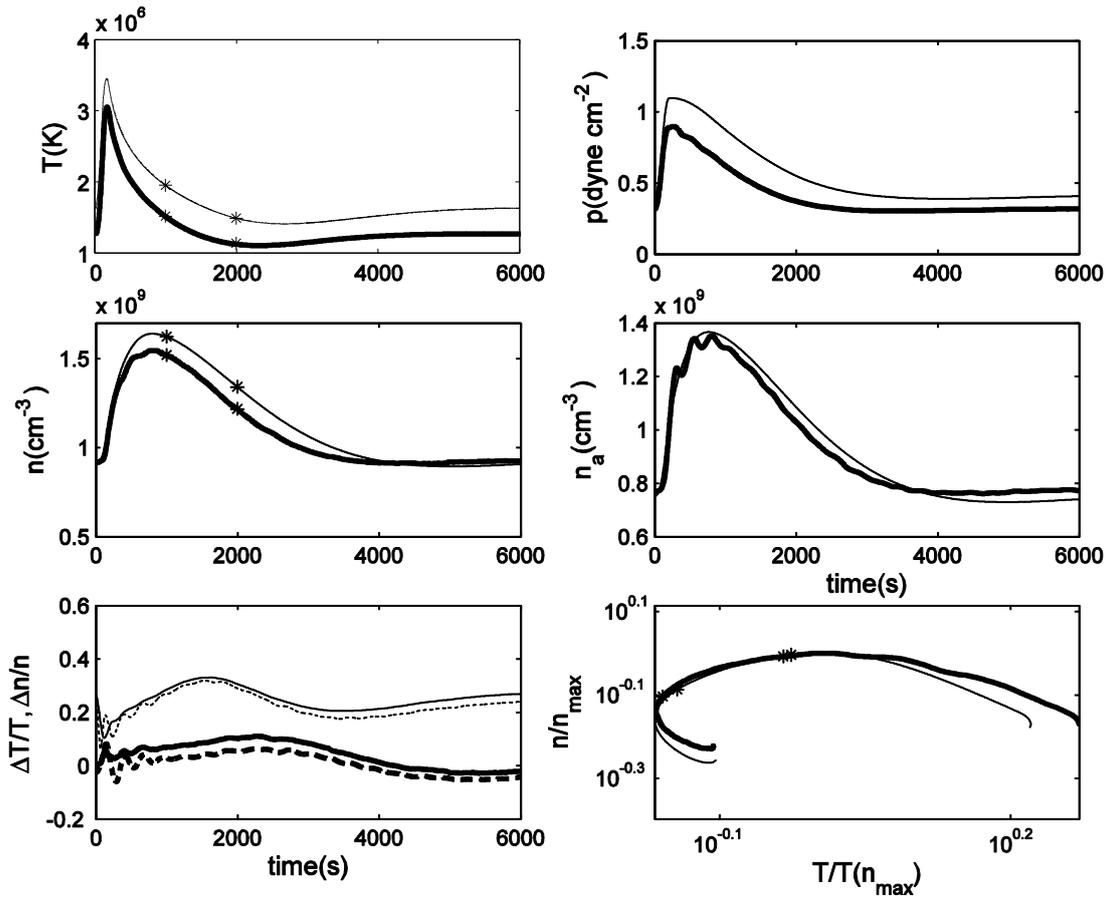

**Figure 7** Case 4. A nanoflare in a dense loop. The format is the same as Figure 5 except we do not show the "EBTEL-1" results.

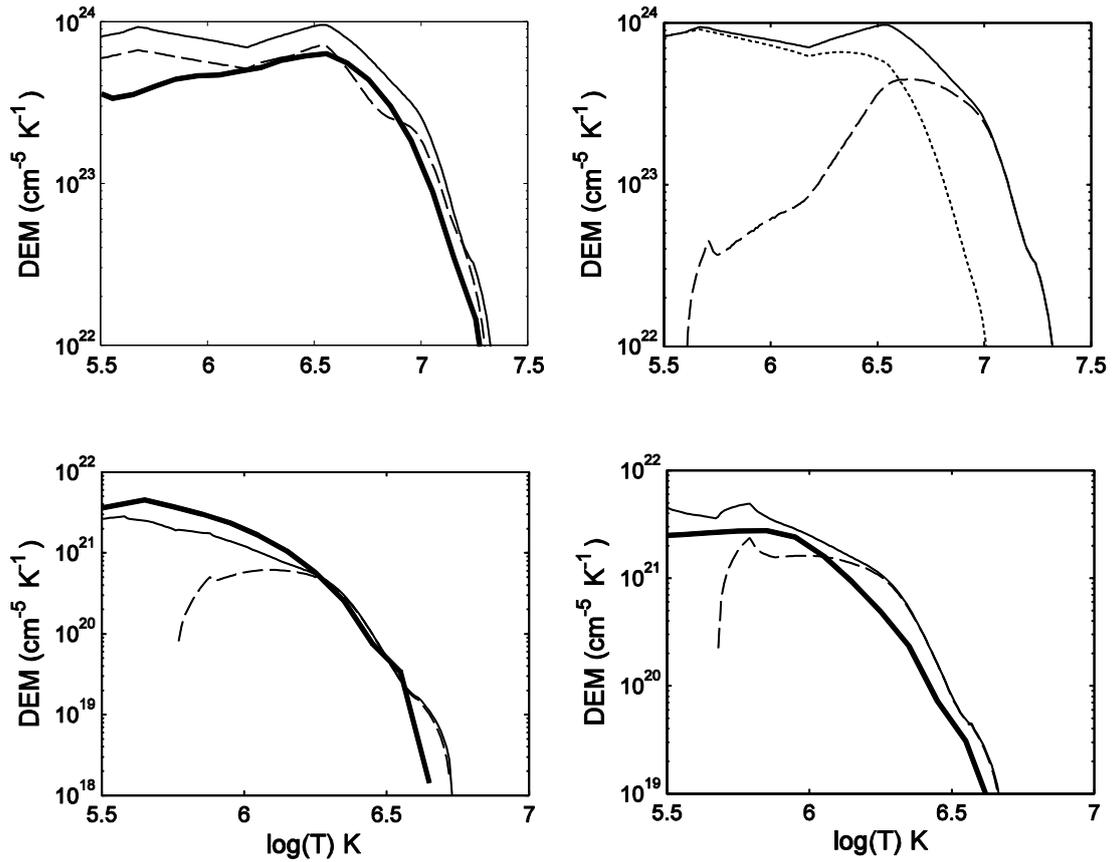

**Figure 8.** The two upper panels show the DEM for Case 3 (flare in a 25 Mm loop). The left panel shows the EBTEL (thin solid line), Hydrad (thick solid line) and EBTEL-1 (dashed line) results. The top right panel shows separate coronal (dashed) and TR (dotted) contributions. The lower panels show the DEM from nanoflares in a long loop (Case 1) and short loop (Case 2) with dashed lines being the coronal component.

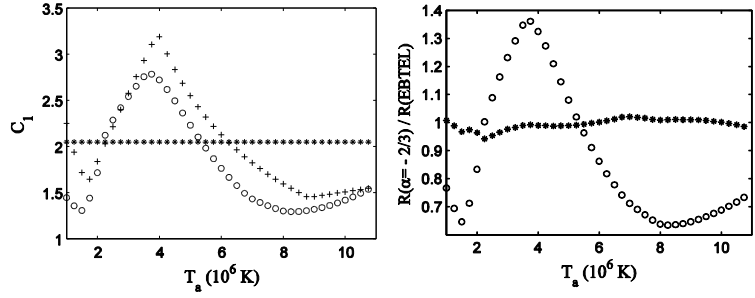

**Figure 9**: $C_1$ as a function of $T_a$ for $L = 5\ 10^9$ cm. There is no gravity. In the left plot stars, circles and plus signs are, respectively, $C_1$ for single power loss function with low temperature correction, $C_1$ for the full EBTEL loss function, and the estimate of $C_1$ in Eq (B1). The right column shows the ratio of radiative losses assuming a single power law and the full EBTEL form in the transition region (stars) and corona (circles).

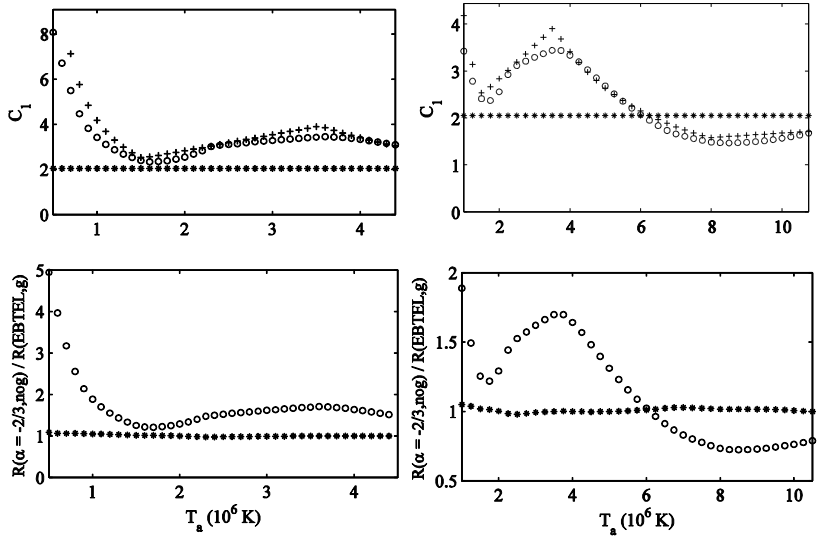

**Figure 10**: The upper row shows $C_1$ as a function of $T_a$ for two temperature ranges and $L = 5\ 10^9$ cm. Stars, circles and plus signs are, respectively, $C_1$ for single power loss function and no gravity, for the EBTEL loss function and gravity, and the estimate of $C_1$ in Eq (B2). The lower row shows the ratio of radiative losses assuming a single power law with low temperature correction and no gravity, and the EBTEL loss function and gravity in the transition region (stars) and corona (circles).